\documentclass[aps,pra,reprint,twocolumns,showpacs]{revtex4-1}
\usepackage{graphicx, amsmath}
\usepackage{amsfonts} % for mathbb
\usepackage[caption=false]{subfig}
\begin{document}

%-------------------------------------------------------------------------------
% Shortcut Commands
%-------------------------------------------------------------------------------

\newcommand{\comb}[2]{{\begin{pmatrix} #1 \\ #2 \end{pmatrix}}}
\newcommand{\braket}[2]{{\left\langle #1 \middle| #2 \right\rangle}}
\newcommand{\bra}[1]{{\left\langle #1 \right|}}
\newcommand{\ket}[1]{{\left| #1 \right\rangle}}
\newcommand{\ketbra}[2]{{\left| #1 \middle\rangle \middle \langle #2 \right|}}

%-------------------------------------------------------------------------------
% Front Matter
%-------------------------------------------------------------------------------

\title{Completeness is Unnecessary for Fast Nonlinear Quantum Search}

\author{David A. Meyer}
	\affiliation{Department of Mathematics, University of California, San Diego, La Jolla, CA 92093-0112}
	\email{dmeyer@math.ucsd.edu}
\author{Thomas G. Wong}
	\affiliation{Department of Physics, University of California, San Diego, La Jolla, CA 92093-0354}
	\altaffiliation{Currently at the University of Latvia}
	\email{twong@lu.lv}

\begin{abstract}
	Although strongly regular graphs and the hypercube are not complete, they are ``sufficiently complete'' such that a randomly walking quantum particle asymptotically searches on them in the same $\Theta(\sqrt{N})$ time as on the complete graph, the latter of which is precisely Grover's algorithm. We show that physically realistic nonlinearities of the form $f(|\psi|^2)\psi$ can speed up search on sufficiently complete graphs, depending on the nonlinearity and graph. Thus nonlinear (quantum) computation can retain its power even when a degree of noncompleteness is introduced.
\end{abstract}

\pacs{03.67.Ac, 05.45.-a, 67.85.Hj, 67.85.Jk}

\maketitle

%-------------------------------------------------------------------------------
% Main Matter
%-------------------------------------------------------------------------------

\section{Introduction}

Although Grover's quantum search algorithm \cite{Grover1996} was originally proposed as a digital algorithm, where the state of the system evolves in discrete-time when acted upon by quantum gates, it can also be formulated as an analog algorithm, where the system evolves in continuous-time by Schr\"odinger's equation with some Hamiltonian. This continuous-time analogue of Grover's algorithm was first proposed by Farhi and Gutmann \cite{FG1998}, and Childs and Goldstone \cite{CG2004} later provided a physical, intuitive way to interpret it as a quantum particle randomly walking on a complete graph with $N$ vertices. Labeling the vertices $\{ | 0 \rangle, | 1 \rangle, \dots, | N - 1 \rangle \}$, the system $| \psi(t) \rangle$ begins in an equal superposition $| s \rangle$ of all the vertices
\[ | \psi(0) \rangle = | s \rangle = \frac{1}{\sqrt{N}} \sum_{i = 0}^{N-1} | i \rangle. \]
Then it evolves by Schr\"odinger's equation with Hamiltonian
\[ H_0 = -\gamma L - | w \rangle \langle w |, \]
where $\gamma$ is the probability amplitude per unit time of the particle jumping to an adjacent vertex (and is inversely proportional to mass), $L$ is the graph Laplacian (an $N$-by-$N$ matrix with $1$ if two vertices are connected, $0$ otherwise, and the negative of the degree of the vertex on the diagonal), and $| w \rangle$ is the ``marked'' vertex that we are searching for. The system evolves in a two-dimensional subspace spanned by $\{ \ket{w}, \ket{s} \}$. When $\gamma$ takes its critical value of $1/N$, the energy eigenvectors are
\begin{equation}
	\label{eq:defn_exact}
	\ket{\psi_{0,1}} = \frac{1}{\sqrt{2}} \sqrt{\frac{\sqrt{N}}{\sqrt{N} \pm 1}} \left( | s \rangle \pm | w \rangle \right),
\end{equation}
with corresponding energy eigenvalues
\begin{equation}
	\label{eq:defn_energies_exact}
	E_{0,1} = -1 \mp \frac{1}{\sqrt{N}},
\end{equation}
So the system evolves from $| s \rangle$ to $| w \rangle$ in time $\pi \sqrt{N} / 2$.

\begin{figure}
\begin{center}
	\subfloat[]{
		\includegraphics[width=1.4in]{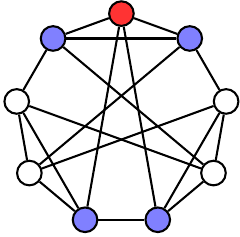}
		\label{fig:srg}
	}
	\quad
	\subfloat[]{
		\includegraphics[width=1.7in]{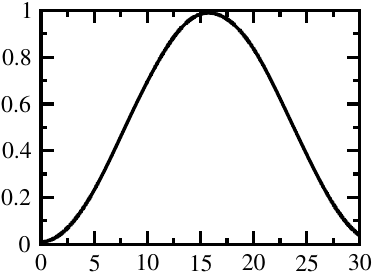}
		\label{fig:prob_time_srg}
	}
	\caption{(a) A strongly regular graph (Paley graph) with parameters (9,4,1,2). The ``marked'' vertex is colored red, vertices ``one away'' (adjacent) are colored blue, and vertices ``two away'' are colored white. (b) Success probability as a function of time for search on a strongly regular graph (Paley graph) with parameters (101,50,24,25).}
\end{center}
\end{figure}

In \cite{JMW2014}, we showed that a randomly walking quantum particle can search for a marked vertex on a strongly regular graph, an example of which is shown in Fig.~\ref{fig:srg}, with the same asymptotic behavior as on the complete graph. An example of this is shown in Fig.~\ref{fig:prob_time_srg} for $N = 101$, and it finds the marked vertex with probability near $1$ at time $\pi \sqrt{101} / 2 \approx 15.786$, as expected. So even though strongly regular graphs are not complete (and even lack global symmetry), they are ``complete enough'' for the search to primarily evolve in its two lowest energy eigenstates, which take the form of \eqref{eq:defn_exact} with corresponding eigenvalues \eqref{eq:defn_energies_exact}, up to terms of order $1/\sqrt{N}$ at the critical $\gamma$. We call such graphs that evolve according to \eqref{eq:defn_exact} and \eqref{eq:defn_energies_exact} with error terms that tend to zero for large $N$ \emph{sufficiently complete}.

\begin{figure}
\begin{center}
	\subfloat[]{
		\includegraphics[width=1.4in]{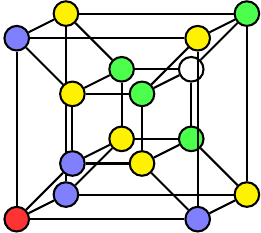}
		\label{fig:hypercube_a}
	}
	\subfloat[]{
		\includegraphics[width=1.7in]{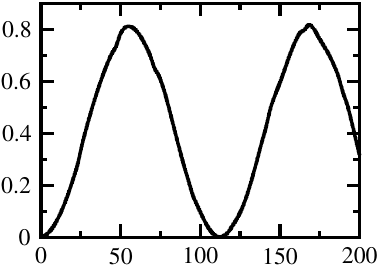}
		\label{fig:hypercube_b}
	}
	\caption{(a) The 4-dimensional hypercube. The ``marked'' vertex is colored red, vertices ``one away'' (adjacent) are blue, vertices ``two away'' are yellow, vertices ``three away'' are green, and the vertex ``four away'' is white. (b) Success probability as a function of time for search on the 10-dimensional hypercube, which has $2^{10} = 1024$ vertices.}
\end{center}
\end{figure}

Strongly regular graphs are not the only sufficiently complete graphs---the hypercube is as well. The hypercube is even ``less complete'' than strongly regular graphs---whereas search on strongly regular graphs evolve in a three-dimensional subspace, search on the $n$-dimensional hypercube evolves in a $(n+1)$-dimensional subspace, which grows with $N = 2^n$. An example of this in four dimensions is shown in Fig.~\ref{fig:hypercube_a}, where vertices that evolve identically are the same color. Nonetheless, the hypercube is complete enough for search to behave like on the complete graph; search on it also primarily evolves in its two lowest eigenstates, which take the form of \eqref{eq:defn_exact} up to terms of order $1/n$ at the critical $\gamma$ \cite{FGGS2000, CDFGGL2002, CG2004}. As an example, the evolution of the success probability for the $10$-dimensional hypercube is shown in Fig.~\ref{fig:hypercube_b}; while $N = 2^{10} = 1024$ is large enough for the runtime to be near $\pi \sqrt{1024} / 2 \approx 50.265$, it is not large enough for the second peak to be near $3\pi \sqrt{1024} / 2 \approx 150.80$, or for the success probability to be near $1$.

As we detailed in \cite{MeyerWong2013, MeyerWong2014}, we can speed up quantum search (\textit{i.e.}, search on the complete graph) by evolving by the nonlinear Schr\"odinger equation
\begin{equation}
	\label{eq:NLSE}
	i \frac{\partial}{\partial t} \psi(\mathbf{r},t) = \big[ \underbrace{H_0 - g f\!\left( |\psi(\mathbf{r},t)|^2 \right)}_H \big] \psi(\mathbf{r},t),
\end{equation}
where $f$ is a real-valued function. Many physically realistic, effective nonlinearities take this form. For example, when $f(p) = p$, \eqref{eq:NLSE} is the Gross-Pitaevskii equation with a cubic nonlinearity that describes Bose-Einstein condensates under certain conditions; when $f(p) = p - p^2$, the cubic-quintic nonlinearity describes nonlinear Kerr media with defocusing corrections; and when $f(p) = \ln p$, the loglinear nonlinearity describes Bose liquids \cite{MeyerWong2013, MeyerWong2014}.

\begin{figure*}
\begin{center}
	\subfloat[]{
		\includegraphics{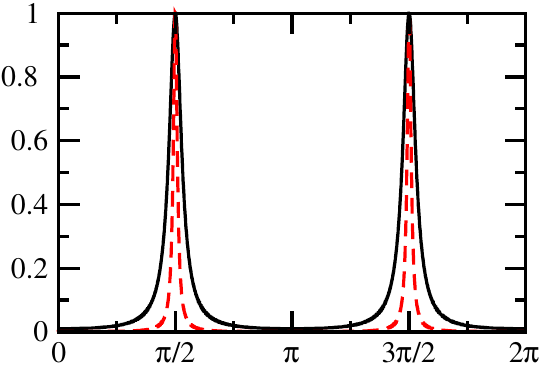}
		\label{fig:complete_cubic}
	}
	\subfloat[]{
		\includegraphics{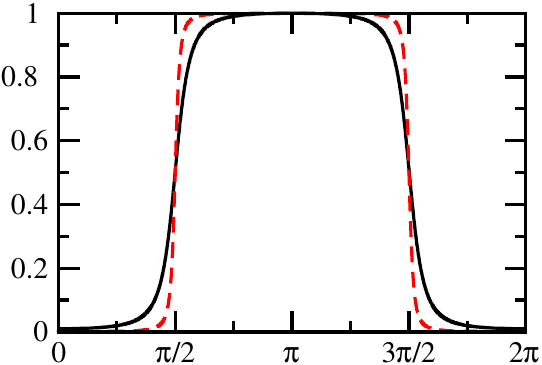}
		\label{fig:complete_cubic-quintic}
	}
	\subfloat[]{
		\includegraphics{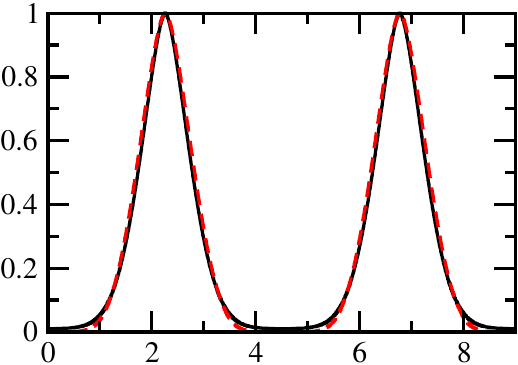}
		\label{fig:complete_loglinear}
	}
	\caption{Success probability as a function of time for search on the complete graph with $N = 100$ and $N = 1000$ vertices (black solid and red dashed curves, respectively) at the critical $\gamma$ with a: (a) cubic nonlinearity and $g = N - 1$, (b) cubic-quintic nonlinearity and $g = N - 1$, and (c) loglinear nonlinearity and $g = \sqrt{N} / \log N$.}
\end{center}
\end{figure*}

By choosing $\gamma$ at its time-varying critical value, the energy eigenvalues remain equal to \eqref{eq:defn_exact}, and $H$ is equal to $H_0$ multiplied by a time-dependent term. This can be interpreted as a rescaling of time, so the nonlinear system evolves identically to the linear algorithm, but with a rescaled time. For example, with the cubic nonlinearity $f(p) = p$, we can choose the nonlinearity coefficient $g$ to achieve a constant-runtime solution, as shown in Fig.~\ref{fig:complete_cubic}. But this improved runtime comes at the novel expense of increasing the time-measurement precision necessary to catch the sudden spike in success probability. Including time-measurement precision as a physical resource, it is better to choose $g$ more modestly to opt for a less aggressive algorithm that runs in $O(N^{1/4})$ time, which is a significant---but not unreasonable---improvement over the $N^{1/2}$ scaling of Grover's algorithm. The cubic-quintic and loglinear nonlinearities, on the other hand, achieve constant-runtime solutions without increasing the time-measurement precision, as shown in Figs.~\ref{fig:complete_cubic-quintic} and \ref{fig:complete_loglinear}.

This raises the question of whether computing with the nonlinear Schr\"odinger equation \eqref{eq:NLSE} speeds up search on sufficiently complete graphs in the same way as it did on the complete graph. In this paper, we show that it can, depending on the precise nonlinearity and graph. This indicates that our model of nonlinear continuous-time (quantum) computation can retain its power even when the underlying graph is not complete.

\section{Linear Search}

We begin by introducing notation to describe linear search on sufficiently complete graphs. First we assume the system exactly evolves with eigenstates and eigenvalues of the form \eqref{eq:defn_exact} and \eqref{eq:defn_energies_exact}, respectively, showing that it finds the marked vertex with probability $1$ in time $\pi \sqrt{N} / 2$. Then we introduce error in the eigenstates and show how it propagates to the success probability and runtime. Since search on strongly regular graphs evolves in a three-dimensional subspace, and search on the hypercube evolves in a $(n+1)$-dimensional subspace, search on a general sufficiently complete graph may evolve in an $M$-dimensional subspace. That is, the $N$ vertices of the graph can be grouped together in $M$ sets $m_i$ of size $|m_i|$, where all vertices in a set evolve identically. Then the equal superpositions of identically evolving vertices
\[ | m_i \rangle = \frac{1}{\sqrt{|m_i|}} \sum_{j \in m_i} | j \rangle \]
form an orthonormal basis $\{ | m_0 \rangle, | m_1 \rangle, \dots, | m_{M-1} \rangle \}$ for the $M$-dimensional subspace.
Without loss of generality, we pick the marked site to be $\ket{m_0} = \ket{w}$, so $|m_0| = 1$. In the case of a strongly regular graph with parameters $(N,k,\lambda,\mu)$, the basis states of the subspace are
\begin{align*}
	\ket{m_0} &= \ket{w} \\
	\ket{m_1} &= \frac{1}{\sqrt{k}} \sum_{(x,w) \in \mathcal{E}} \ket{x} \\
	\ket{m_2} &= \frac{1}{\sqrt{N-k-1}} \sum_{(x,w) \not\in \mathcal{E}} \ket{x},
\end{align*}
which correspond to the marked vertex, vertices adjacent to the marked vertex, and vertices not adjacent to the marked vertex. For the $n$-dimensional hypercube, we first label each of the $N = 2^n$ vertices with an $n$-bit string $\ket{z_1 \dots z_n}$. Without loss of generality, we choose the marked vertex to be the string of all zeros: $\ket{w} = \ket{0 \dots 0}$. Then the vertices ``one away'' are bit strings with a single one (\textit{i.e.}, with Hamming weight 1), the vertices ``two away'' are bit strings with two ones (\textit{i.e.}, with Hamming weight 2), and so forth. Taking the superposition of vertices equally far from the marked vertex, we get
\[ \ket{m_k} = \comb{n}{k}^{-1/2} \sum_{z_1 + \dots + z_n = k} \ket{z_1 \dots z_n}. \]
Then the set $\{ \ket{m_k} : k = 0, 1, \dots, n \}$ is an orthonormal basis for the $(n+1)$-dimensional subspace.

In this $\{ | m_0 \rangle, | m_1 \rangle, \dots, | m_{M-1} \rangle \}$ subspace, the state $| \psi(t) \rangle$ of the system can be written as a linear combination of the basis states:
\[ | \psi(t) \rangle = \sum_{i=0}^{M-1} c_i(t) | m_i \rangle. \]
We assume that the system evolves in its two lowest energy eigenstates, having the form \eqref{eq:defn_exact}. Then the amplitudes are
\begin{align*}
	c_i(t) 
		&= \braket{m_i}{\psi(t)} = \langle m_i | e^{-iHt} | s \rangle \\
		&= \braket{m_i}{\psi_0} \braket{\psi_0}{s} e^{-iE_0t} + \braket{m_i}{\psi_1} \braket{\psi_1}{s} e^{-iE_1t}. 
\end{align*}
Using \eqref{eq:defn_exact}, the inner products are
\begin{align*}
	\braket{\psi_{0,1}}{s} 
		&= \frac{1}{\sqrt{2}} \sqrt{\frac{\sqrt{N}}{\sqrt{N} \pm 1}} \left( 1 \pm \frac{1}{\sqrt{N}} \right) \\
		&= \frac{1}{\sqrt{2}} \sqrt{\frac{\sqrt{N}}{\sqrt{N} \pm 1}} \frac{\sqrt{N} \pm 1}{\sqrt{N}}
\end{align*}
and
\[ \braket{m_i}{\psi_{0,1}} = \frac{1}{\sqrt{2}} \sqrt{\frac{\sqrt{N}}{\sqrt{N} \pm 1}} \left( \frac{\sqrt{|m_i|}}{\sqrt{N}} \pm \delta_{i0} \right), \]
where $\delta_{i0} = 1$ when $i = 0$ and $0$ otherwise is the Kronecker delta.
Then the amplitudes are
\[ c_i(t) = \begin{cases}
	e^{-it} \left[ \frac{1}{\sqrt{N}} \cos \left( \frac{t}{\sqrt{N}} \right) + i \sin \left( \frac{t}{\sqrt{N}} \right) \right], & i = 0 \\
	e^{-it} \frac{\sqrt{|m_i|}}{\sqrt{N}} \cos \left( \frac{t}{\sqrt{N}} \right), & i \ne 0 \\
\end{cases}. \]
Squaring them, the probabilities are
\[
	| c_i(t) |^2 = 
	\begin{cases}
		\frac{1}{N} \cos^2 \left( \frac{t}{\sqrt{N}} \right) + \sin^2 \left( \frac{t}{\sqrt{N}} \right), & i = 0 \\
		\frac{|m_i|}{N} \cos^2 \left( \frac{t}{\sqrt{N}} \right), & i \ne 0 \\
	\end{cases}.
\]
This reveals that the success probability $|c_0|^2$ reaches $1$ at time $\pi \sqrt{N} / 2$.

Now, to introduce error, a sufficiently complete graph has eigenstates and eigenenergies of the form \eqref{eq:defn_exact} and \eqref{eq:defn_energies_exact}, respectively, both up to terms of some order $\epsilon$ that tends to zero for large $N$:
\begin{equation}
	\label{eq:defn}
	\ket{\psi_{0,1}} = \frac{1}{\sqrt{2}} \sqrt{\frac{\sqrt{N}}{\sqrt{N} \pm 1}} \left( | s \rangle \pm | w \rangle \right) \left( 1 + O(\epsilon) \right),
\end{equation}
\begin{equation}
	\label{eq:defn_energies}
	E_{0,1} = \left( -1 \mp \frac{1}{\sqrt{N}} \right) \left( 1 + O(\epsilon) \right).
\end{equation}
For strongly regular graphs, $\epsilon = 1/\sqrt{N}$, and for the $n$-dimensional hypercube, $\epsilon = 1/n$. Propagating these errors through the previous calculations, we get probabilities
\begin{align*}
	\left| c_0(t) \right|^2 = \bigg\{ &\frac{1}{N} \cos^2 \left[ \frac{t}{\sqrt{N}} \left( 1 + O(\epsilon) \right) \right] \\ &+ \sin^2 \left[ \frac{t}{\sqrt{N}} \left( 1 + O(\epsilon) \right) \right] \bigg\} (1 + O(\epsilon)),
\end{align*}
and
\[ \left| c_{i \ne 0}(t) \right|^2 = \frac{|m_i|}{N} \cos^2 \left[ \frac{t}{\sqrt{N}} \left( 1 + O(\epsilon) \right) \right] (1 + O(\epsilon)). \]
Thus the system evolves from $\ket{s}$ to $\ket{w}$ in time $\pi \sqrt{N} / 2$ for large $N$, as expected.

\section{Nonlinear Search}

For the nonlinear algorithm, we subtract from $H_0$ an additional nonlinear ``self-potential'' $V(t) = g f(|\psi(\mathbf{r},t)|^2)$, where $f$ is a real-valued function, so that the system evolves according to the nonlinear Schr\"odinger equation \eqref{eq:NLSE}. Then for positive $g$ this speeds up the buildup of probability amplitude. In the computational basis, the self-potential is
\begin{equation}
	\label{eq:v}
	V(t) = g \sum_{i=0}^{N-1} f\!\left( \left| \langle i | \psi \rangle \right|^2 \right) | i \rangle \langle i |.
\end{equation}
Even with this nonlinearity, the system still evolves in the $M$-dimensional subspace spanned by $\{ | m_0 \rangle, | m_1 \rangle, \dots, | m_{M-1} \rangle \}$. In this $M$-dimensional subspace, the self-potential \eqref{eq:v} has off-diagonal elements equal to zero. Its diagonal terms are
\begin{align*}
	\left\langle m_i \middle| V(t) \middle| m_i \right\rangle 
		&= g \sum_{j = 0}^{N-1} f\!\left( \left| \langle j | \psi \rangle \right|^2 \right) \langle m_i | j \rangle \langle j | m_i \rangle \\
		&= g \frac{1}{|m_i|} \sum_{j \in m_i} f\!\left( \left| \langle j | \psi \rangle \right|^2 \right) \\
		&= g \frac{1}{|m_i|} |m_i| f\!\left( \frac{|c_i|^2}{|m_i|} \right) \\
		&= g f\!\left( \frac{|c_i|^2}{|m_i|} \right).
\end{align*}
For ease of notation, let us define
\[ f_i = f\!\left( \frac{|c_i|^2}{|m_i|} \right). \]
Then in the $M$-dimensional subspace, the nonlinearity is
\[ V(t) = g \begin{pmatrix}
	f_0 & 0 & \cdots & 0 \\
	0 & f_1 & \cdots & \vdots \\
	\vdots & \vdots & \ddots & 0 \\
	0 & \cdots & 0 & f_{M-1}
\end{pmatrix}. \]

In the $M$-dimensional subspace, the equation of motion is governed by
\begin{align*}
	H 
	&= -\gamma L - \ketbra{m_0}{m_0} - g \sum_{i = 0}^{M-1} f_i \ketbra{m_i}{m_i} \\
	&= -\gamma L - \left( 1 + g f_0 \right) \ketbra{m_0}{m_0} - g \sum_{i = 1}^{M-1} f_i \ketbra{m_i}{m_i} \\
	&= -\gamma L - \left( 1 + g f_0  - g f_1 \right) \ketbra{m_0}{m_0} - g f_1 \mathbb{I} \\
	&\quad\quad - g \sum_{i = 2}^{M-1} \left( f_i - f_1 \right) \ketbra{m_i}{m_i} .
\end{align*}
The term proportional to the identity matrix can be dropped since it is a rescaling of energy (or an overall phase), which is unobservable. We want to show that there exists a critical $\gamma$ that causes the nonlinear evolution to follow the same path as the linear evolution for large $N$, but with rescaled time $\tau$. That is, we want to show that $g(f_{i \ge 2} - f_1)$ can be dropped in comparison to $1 + g(f_0 - f_1)$ when we use the linear evolution, but with $t \to \tau$. Assume for the moment that we can do this. Then we have
\[ H = -\gamma L - \left( 1 + g f_0 - g f_1 \right) \ketbra{m_0}{m_0} \]
for large $N$. Then the critical $\gamma$ for the nonlinear algorithm is
\[ \gamma_c = \gamma_L \left( 1 + g f_0 - g f_1 \right), \]
where $\gamma_L$ is the linear algorithm's critical $\gamma$. At $\gamma_c$, we have
\[ H = \left( 1 + g f_0 - g f_1 \right) \left( -\gamma_L L - \ketbra{m_0}{m_0} \right), \]
which is the linear Hamiltonian at its critical $\gamma$ with a rescaled factor that depends on $f_0 - f_1$. Thus the critical $\gamma$ causes the nonlinear algorithm to evolve identically to the linear algorithm, but with rescaled time. For large $N$, the leading order behavior of $f_0 - f_1$ is
\begin{align*}
	f_0 - f_1 
		&= f\!\left( \frac{1}{N} \cos^2 \left( \frac{\tau}{\sqrt{N}} \right) + \sin^2 \left( \frac{\tau}{\sqrt{N}} \right) \right) \\
		&\quad- f\!\left( \frac{1}{N} \cos^2 \left( \frac{\tau}{\sqrt{N}} \right) \right)
\end{align*}
which is the same for all sufficiently complete graphs that can be sped up by the nonlinearity, including complete graphs, and so we expect the nonlinearity to speed them up the same way.

\begin{figure*}
\begin{center}
	\subfloat[]{
		\includegraphics{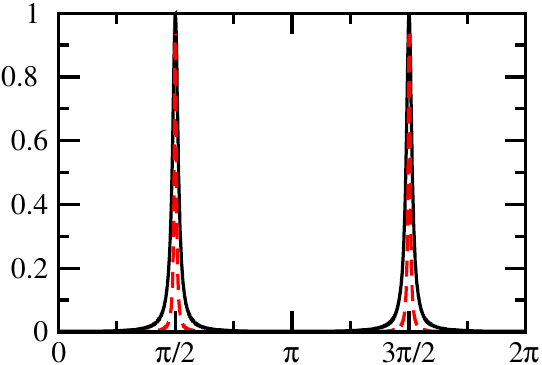}
		\label{fig:srg_cubic}
	}
	\subfloat[]{
		\includegraphics{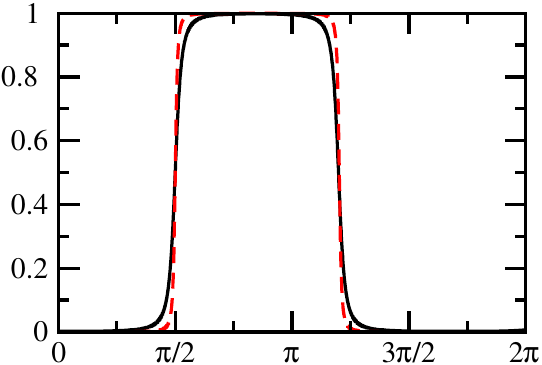}
		\label{fig:srg_cubic-quintic}
	}
	\subfloat[]{
		\includegraphics{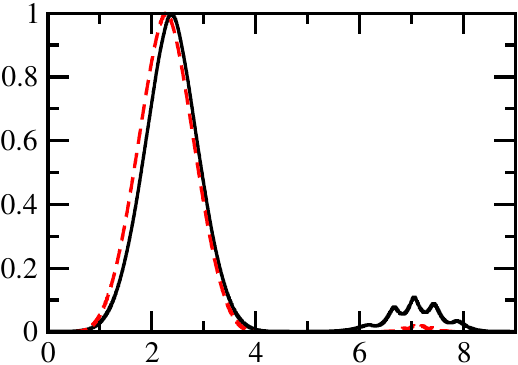}
		\label{fig:srg_loglinear}
	}
	\caption{Success probability as a function of time for search on strongly regular graphs with parameters $(N, k, \lambda, \mu)$ = (509,254,126,127) and (4001,2000,999,1000) (black solid and red dashed curves, respectively) at the critical $\gamma$ with a: (a) cubic nonlinearity and $g = N - 1$, (b) cubic-quintic nonlinearity and $g = N - 1$, and (c) loglinear nonlinearity and $g = \sqrt{N}/\log N$.}
\end{center}
\end{figure*}

\begin{figure*}
\begin{center}
	\subfloat[]{
		\includegraphics{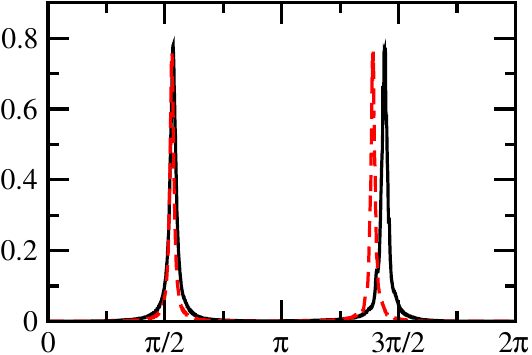}
		\label{fig:hypercube_cubic}
	}
	\subfloat[]{
		\includegraphics{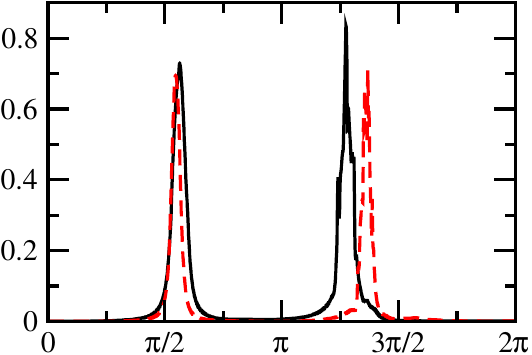}
		\label{fig:hypercube_cubic-quintic}
	}
	\subfloat[]{
		\includegraphics{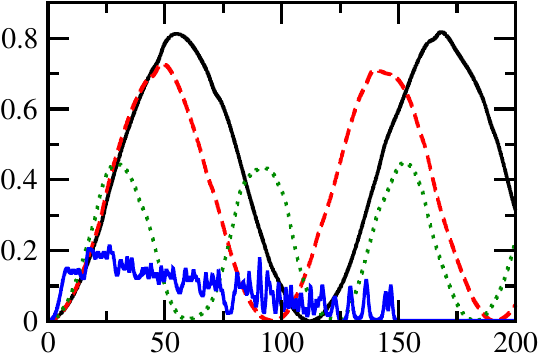}
		\label{fig:hypercube_loglinear}
	}
	\caption{Success probability as a function of time for search on the $10$-dimensional and $11$-dimensional hypercubes (black solid and red dashed curves, respectively) at the critical $\gamma$ with a: (a) cubic nonlinearity and $g = N - 1$ and (b) cubic-quintic nonlinearity and $g = N - 1$. Subfigure (c) is search on the $10$-dimensional hypercube with $g = 0$ (linear), $g = 0.01$, $g = 0.05$, and $g = 0.5$ (black solid, red dashed, green dotted, and wildly varying blue solid curves, respectively).}
\end{center}
\end{figure*}

Now let us prove that such a critical $\gamma$ exists, namely that we can drop $g(f_{i \ge 2} - f_1)$ compared to $1 + g(f_0 - f_1)$ with rescaled time $t \to \tau$ in Eqs.~\eqref{eq:defn} and \eqref{eq:defn_energies}, beginning with the cubic nonlinearity $f(p) = p$. At later time, the sine term in $f_0$ clearly dominates both $f_{i \ge 2}$ and $f_1$. To determine what happens at early time, we start with
\[ f_{i \ne 0} = \frac{1}{N} \cos^2 \left[ \frac{\tau}{\sqrt{N}} \left( 1 + O(\epsilon) \right) \right] (1 + O(\epsilon)). \]
Note that $\cos^2(a(1+x)) \approx \cos^2(a) - 2a \cos(a) \sin(a) x$ for small $x$. Then we have
\begin{align*}
	f_{i \ne 0} &\approx \frac{1}{N} \Big[ \cos^2 \left( \frac{\tau}{\sqrt{N}} \right) \\ &- 2 \frac{\tau}{\sqrt{N}} \cos \left( \frac{\tau}{\sqrt{N}} \right) \sin \left( \frac{\tau}{\sqrt{N}} \right) O(\epsilon) \Big] \left( 1 + O(\epsilon) \right).
\end{align*}
For short time, $\cos(\tau/\sqrt{N}) \approx 1$ and $\sin(\tau/\sqrt{N}) \approx \tau/\sqrt{N}$. Also absorbing $-2$ into the big-$O$, we get
\[ f_{i \ne 0} \approx \frac{1}{N} \left[ \cos^2 \left( \frac{\tau}{\sqrt{N}} \right) + \left( \frac{\tau}{\sqrt{N}} \right)^2 O(\epsilon) \right] \left( 1 + O(\epsilon) \right). \]
Multiplying out the terms, using $\cos(\tau/\sqrt{N})~\approx~1$, and taking $\tau = O(1)$ we get,
\[ f_{i \ne 0} \approx \frac{1}{N} \left[ \cos^2 \left( \frac{\tau}{\sqrt{N}} \right) + O(\epsilon) \right]. \]
Then the terms we want to drop are, at short time,
\[ g(f_{i \ge 2} - f_1) \approx O\left(\frac{g \epsilon}{N}\right). \]
This can be dropped compared to $1+g(f_0 - f_1)$ when $g = \Theta(N)$ for both strongly regular graphs and the hypercube, so the constant-runtime algorithm exists for both, as shown in Figs.~\ref{fig:srg_cubic} and \ref{fig:hypercube_cubic}.

These arguments persist when $f(p) = p^q$---the sine term in $f_0$ dominates $f_{i \ne 0}$ at later time, and we have at early time
\[ f_{i \ne 0} \approx \frac{1}{N^q} \left[ \cos^{2q} \left( \frac{\tau}{\sqrt{N}} \right) + O(\epsilon) \right]. \]
Using this for the cubic-quintic nonlinearity $f(p) = p - p^2$, we have
\begin{align*}
	f_{i \ne 0} 
		&= \frac{1}{N} \left[ \cos^{2} \left( \frac{\tau}{\sqrt{N}} \right) + O(\epsilon) \right] \\
		&\quad - \frac{1}{N^2} \left[ \cos^{4} \left( \frac{\tau}{\sqrt{N}} \right) + O(\epsilon) \right].
\end{align*}
So the terms we want to drop are
\[ g(f_{i \ge 2} - f_1) = O\left(\frac{g \epsilon}{N}\right), \]
which can be dropped compared to $1+g(f_0 - f_1)$ when $g = \Theta(N)$ for both strongly regular graphs and the hypercube. So the constant-runtime algorithm exists for both, as shown in Figs.~\ref{fig:srg_cubic-quintic} and Fig.~\ref{fig:hypercube_cubic-quintic}. Note that the peak in success probability is not as wide as for the complete graph in Fig.~\ref{fig:complete_cubic-quintic}; this is expected because the width broadens as the success probability approaches 1. That is, as it approaches 1, $f_0$ goes to zero because of cancellation between the cubic and quintic components, as does $f_1$ because the failure probability goes to zero. Then the rescaling of time ceases, causing a broad peak. The error with which the success probability approaches $1$ decreases as $N$ increases, so the peak is wider for larger $N$, as seen for strongly regular graphs in Fig.~\ref{fig:srg_cubic-quintic}. For the hypercube in Fig.~\ref{fig:hypercube_cubic-quintic}, the success probability is too low, so we do not expect the peak to broaden at all.

\begin{figure}
\begin{center}
	\includegraphics{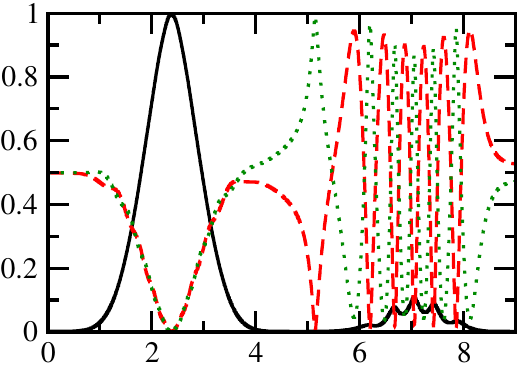}
	\caption{\label{fig:srg_evolution_loglinear}Evolution of search with a loglinear nonlinearity and $g = \sqrt{N}/\log N$ on a strongly regular graph with parameters $(N, k, \lambda, \mu)$ = (509,254,126,127). The black solid curve is $|c_0(t)|^2$, the red dashed curve is $|c_1(t)|^2$, and the green dotted curve is $|c_2(t)|^2$; they correspond to the marked vertex, vertices adjacent to the marked vertex, and vertices not adjacent to the marked vertex, respectively.}
\end{center}
\end{figure}

Now consider the loglinear nonlinearity $f(p) = \log p$. Using the short time approximations from before and $\log a - \log b = \log(a/b)$, we have
\[ f_{i \ge 2} - f_1 = \log \left( \frac{\cos^2 \left( \frac{\tau}{\sqrt{N}} \right) + O(\epsilon)}{\cos^2 \left( \frac{\tau}{\sqrt{N}} \right) + O(\epsilon)} \right). \]
Note that $1/(a+x) \approx 1/a + (1/a^2) x$ for small $x$. Then we get
\begin{align*}
	f_{i \ge 2} - f_1 
		&\approx \log \left(\rule{0cm}{0.7cm}\right. \left[ \cos^2 \left( \frac{\tau}{\sqrt{N}} \right) + O(\epsilon) \right] \\
		&\quad \times \left[ \frac{1}{\cos^2 \left( \frac{\tau}{\sqrt{N}} \right)} - \frac{1}{\cos^4 \left( \frac{\tau}{\sqrt{N}} \right)} O(\epsilon) \right] \left.\rule{0cm}{0.7cm}\right).
\end{align*}
Multiplying this out and absorbing minus signs into the big-$O$'s,
\[ f_{i \ge 2} - f_1 \approx \log \left( 1 + \frac{O(\epsilon)}{\cos^2 \left( \frac{\tau}{\sqrt{N}} \right)}  + \frac{O(\epsilon^2)}{\cos^4 \left( \frac{\tau}{\sqrt{N}} \right)} \right). \]
For short time, $\cos(\tau/\sqrt{N}) \approx 1$, so
\[ f_{i \ge 2} - f_1 \approx \log \left( 1 + O(\epsilon) \right). \]
Since $\log(1+x) \approx x$ for small $x$, the terms we want to drop are
\[ g(f_{i \ge 2} - f_1) \approx O(g\epsilon). \]
The constant-runtime algorithm requires $g = \Theta(\sqrt{N}/\log N)$, so we can drop this compared to $1+g(f_0 - f_1)$ for strongly regular graphs, as shown in Fig.~\ref{fig:srg_loglinear}. The second ``peak'' is strange because of numerical error; the derivative of $\log x$ at $x = 0$ is nonzero, which makes the nonlinearity highly susceptible to noise, as shown in Fig.~\ref{fig:srg_evolution_loglinear}, where the evolution begins to vary wildly shortly after the first peak. For the hypercube, the constant-runtime algorithm does not exist. In fact, $g$ must scale less than $n = \log N$, which is not significant enough for the loglinear nonlinearity to provide substantial speedup, as shown in Fig.~\ref{fig:hypercube_loglinear}.

So whether a sufficiently complete graph can be sped up by the nonlinear Schr\"odinger equation depends on the graph and the nonlinearity. Nonetheless, we have shown that even with a degree of noncompleteness, some nonlinearities speed up search on certain sufficiently complete graphs in the same way as on the complete graph for large $N$. Thus completeness is not a requirement for fast nonlinear (quantum) search.

\begin{acknowledgments}
	This work was partially supported by the Defense Advanced Research Projects Agency as part of the Quantum Entanglement Science and Technology program under grant N66001-09-1-2025, the Air Force Office of Scientific Research as part of the Transformational Computing in Aerospace Science and Engineering Initiative under grant FA9550-12-1-0046, and the Achievement Awards for College Scientists Foundation.
\end{acknowledgments}

%-------------------------------------------------------------------------------
% References.
%-------------------------------------------------------------------------------

\bibliography{refs}

\end{document}